\documentclass[prc,aps,floatfix,groupedaddress,amsmath,amssymb,twocolumn]{revtex4}

\usepackage{graphicx}
\usepackage{dcolumn}
\usepackage{bm}
\usepackage{color}

\usepackage[section]{placeins}
\usepackage{graphicx,epsfig}
\usepackage{amsmath}
\usepackage{amsfonts}
\usepackage{braket}
\usepackage{fancyhdr}
\usepackage{hyperref}
\usepackage[utf8]{inputenc}
\usepackage[T1]{fontenc}
\usepackage{amsthm}
\usepackage{hhline}
\usepackage{multirow}
\usepackage[figuresleft]{rotating}
\usepackage{gensymb}
\usepackage{mathptm}

\begin{document}

\title{Femtoscopic correlation function for the $T_{cc}(3875)^+$ state}

\author{I. Vida\~na$^1$, A. Feijoo$^2$, M. Albaladejo$^2$, J. Nieves$^2$ and E. Oset$^{2,3,4}$}
\affiliation{$^1$Istituto Nazionale di Fisica Nucleare, Sezione di Catania, Dipartimento di Fisica ``Ettore Majorana'', Universit\`a di Catania, Via Santa Sofia 64, I-95123 Catania, Italy}
\affiliation{$^2$Instituto de F\'{i}sica Corpuscular, Centro Mixto Universidad de Valencia-CSIC, Institutos de Investigaci\'{o}n de Paterna, Aptdo. 22085, E-46071 Valencia, Spain}
\affiliation{$^3$Departamento de F\'{i}sica Te\'{o}rica and IFIC, Centro Mixto Universidad de Valencia-CSIC, Institutos de Investigaci\'{on } de Paterna, Aptdo. 22085, E-46071 Valencia, Spain}
\address{$^4$Department of Physics, Guangxi Normal University, Guilin 541004, China}


\begin{abstract}
We have conducted a study of the femtoscopic correlation functions for the $D^0D^{*+}$ and $D^+D^{*0}$ channels that build the $T_{cc}$ state. We develop a formalism that allows us to factorize the scattering amplitudes outside the integrals in the formulas, and the integrals involve the range of the strong interaction explicitly.  For a source of size of 1 fm, we find values for the correlation functions of the $D^0 D^{*+}$ and  $D^+D^{*0}$  channels at the origin around 30 and 2.5, respectively, and we see these observables converging to unity already for relative momenta of the order of 200 MeV. We conduct tests to see the relevance of the different contributions to the correlation function and find that it mostly provides information on the scattering length, since the presence of the source function in the correlation function introduces an effective cut in the loop integrals that makes them quite insensitive to the range of the interaction. 
\end{abstract}

\maketitle

{\it Keywords:} Femtoscopy, $T_{cc}$


\section{Introduction}

The study of molecular states has long history (see {\it e.g.} Refs.\ \cite{Oller00,Guo18,Dong21,Guo21,Dong21b}) implying that many known mesonic states would actually contain four quarks (two quarks and two antiquarks), and baryonic states five quarks (four quarks and one antiquark), hence diverting from the $q\bar q$ standard nature associated to the mesons and $qqq$ for baryons. The big experimental boost to this idea came from the discovery of the $P_c$ and $P_{cs}$ pentaquark states by LHCb~\cite{Aaij15,Aaij16,Aaij19,Aaij21,Aaij22b}. While these states still contain quantum numbers compatible with a three quark structure, their large mass precludes them to correspond to excited non charmed $qqq$ baryons. The cleanest cases, however, were found with the discovery of mesonic states containing two open quarks of different nature. One of them was the $X_0(2900)$ [$T_{cs}(2900)$] \cite{Aaij20,Aaij20b}, found in the $D^-K^+$ mass distribution in the $B^+\rightarrow D^+D^-K^+$ decay, with $\bar c\bar s$-quark content. The other one is the $T_{cc}(3875)^+$ state reported in \cite{Aaij22,Aaij22c}, observed in the $D^0D^0\pi^+$ mass distribution, which has two open charmed quarks. The observation of this extremely narrow state, which has a small binding with respect to the $D^0D^{*+}$ threshold $m_\text{th}=3875.09$ MeV, has spurred lots of interest in the hadronic community. Its mass and width are $m_\text{th}+\delta m_\text{exp}$, with $\delta m_\text{exp}=-360\pm 40^{+4}_{-0}$ keV, and  $\Gamma=48\pm 2^{+0}_{-14}$ keV, respectively~\cite{Aaij22c}. 

The proximity of the mass of this exotic state to the $D^0D^{*+}$ and $D ^+D^{*0}$ thresholds has given support to the molecular picture in these channels~\cite{Dong21,Feijoo21,Ling22,Fleming21,Ren22,Chen22,Albaladejo22,Du22,Baru:2021ldu,Santowsky22,Deng22,Ke22,Agaev22,Kamiya22,Meng22,Abreu22,Chen22b,Albaladejo:2022sux}, although tetraquark schemes have also been invoked, even prior to the $T_{cc}$ discovery \cite{Ballot83,Zouzou86}. Even then, the extreme proximity of the states to the $D^0D^{*+}$ and $D^+D^{*0}$ thresholds makes it unavoidable to include these degrees of freedom in any analysis of the experimental data as shown in \cite{Guo21}.

Additional independent information on this state would be most welcome to further learn about its properties and nature. Hence, the information that can be provided by the femtoscopic correlation functions with the observation of the $D^0D^{*+}$ and $D ^+D^{*0}$ channels in heavy ion collisions, should be most valuable. The idea of using the correlation functions to learn about hadron interactions is relatively new, but there are already many theoretical \cite{Morita15,Onishi16,Morita16,Hatsuda17,Mihaylov18,Haidenbauer19,Morita20,Kamiya20,Kamiya22b} and experimental \cite{Adamczyk15,Acharya17,Adam19,Acharya19,Acharya19b,Acharya19c,Acharya20,Acharya20b,Acharya20c,Acharya21,Acharya21b,Fabietti21} studies on the subject in the strangeness sector, and work on the charm sector has already started (see {\it e.g.,} the ALICE collaboration analysis on the $Dp$ channel \cite{Acharya22}).
 
In this work, we take the case of the $T_{cc}(3875)^+$ exotic state and evaluate the correlation function for the channels $D^0D^{*+}$ and $D ^+D^{*0}$. Work on this issue has already been done in \cite{Kamiya22} using coordinate space wave functions and potentials. Here we divert from this picture and use a momentum space formalism, as has also been used recently in \cite{Wei23} for the correlations generated from the $DK$ interaction. We rely upon the framework used in \cite{Feijoo21}, where the local hidden gauge approach
\cite{Bando88,Harada03,Meissner88,Nagahiro09}, exchanging vector mesons as a source of the interaction, has been used. The description of the $T_{cc}$ was done using the two coupled channels $D^0D^{*+}$ and $D ^+D^{*0}$, and a bound state was obtained which has essentially\footnote{Isospin symmetry is not exact due to the small difference between the $D^0D^{*+}$ and $D ^+D^{*0}$ thresholds, which cannot be neglected because the remarkable proximity of the mass of the $T_{cc}$ to the $D ^+D^{*0}$ threshold.   }  isospin $I=0$, as also determined in the experimental work \cite{Aaij22c}. We compare our results with those obtained in \cite{Kamiya22} and also make some comments concerning the formalism used in \cite{Wei23}. 
We take advantage to show that the correlation functions in this case provide mostly information on the scattering length and very little on the range of the interaction. Indeed, we discuss how the range of the interaction affects the correlation functions, concluding that, due to the extension of the source, the results are rather insensitive to the range of the interaction.

The manuscript is organized as follows. A description of the correlation function formalism is presented in Sec.~\ref{sec:form} followed by a brief summary of the study of the $T_{cc}$ state of Ref.\ \cite{Feijoo21} in Sec.~\ref{sec:tcc}. The results for the correlation functions of the $D^0D^{*+}$ and $D^+D^{*0}$ pairs and the different tests are discussed in Sec.~\ref{sec:results}. In Sec.~\ref{sec:compar}, we obtain the correlation functions using as an input the $T$-matrix of Ref.~\cite{Albaladejo22} and compare with those obtained in Sec.~\ref{sec:results}. Finally, the main conclusions are presented in Sec.\ \ref{sec:conclusions}.

\section{Formalism for the correlation function}
\label{sec:form}

The correlation function is defined as the ratio of the probability to find the two interacting particles and the product of the probabilities
to find each individual particle \cite{Lisa05}. Within certain approximations, it can be written as \cite{Onishi16}
\begin{equation}
C(\vec p\,)=\int d^3\vec r S_{12}(\vec r\,)|\Psi(\vec r,\vec p\,)|^2
\label{eq:cf}
\end{equation}
where $S_{12}(\vec r\, )$ is the so-called source function, usually parametrized as a Gaussian normalized to unity,
\begin{equation}
S_{12}(\vec r\, ) = S_{12}(r) = \frac{1}{(\sqrt{4\pi})^3 R^3}\mbox{exp}\left(-\frac{r^2}{4R^2}\right) \ .
\label{sec:source}
\end{equation}
Here $R$ is the size of the source and it takes values in the range $1-5$ fm depending on the kind of scattering used to produce the particles. Typically,  $R\simeq 1$ fm for proton-proton collisions and $R\simeq 5$ fm in the case of heavy ion collisions. In addition, $\vec p$ in
Eq.~\eqref{eq:cf} is the relative momentum of the two particles, i.e., the momentum of each particle in the center-of-mass (c.m.) frame of the
two particles, while $\Psi(\vec r,\vec p\,)$ is the c.m. outgoing wave function of the two particles.\footnote{ In the case of dealing with complex optical potentials, it is useful to recall that  in Eq.~\eqref{eq:cf} one should use $\Psi^{(-)}(\vec r,\vec p\,)$ and that
$\left[\Psi^{(-)}(\vec r,\vec p\,)\right]^*$, as a block, corresponds to an incoming solution of the Schr\"{o}dinger equation \cite{Nieves93} (see also appendix A of Ref.\ \cite{Kamiya22} reaching the same conclusion).} We can thus write the Schr\"{o}dinger equation for the interacting wave function $\Psi$ of the pair with $H=H_0+V$ ($H_0$ the kinetic energy)
\begin{equation}
H\Psi=(H_0+V)\Psi=E\Psi \Rightarrow (E-H_0)\Psi=V\Psi
\label{eq:schroed}
\end{equation}
while the free wave function $\Phi$, when $V\equiv 0$, for the same energy $E$ satisfies,
\begin{equation}
(E-H_0)\Phi=0 \ .
\label{eq:schroed_free}
\end{equation} 
Subtracting Eqs.~\eqref{eq:schroed} and \eqref{eq:schroed_free} we obtain
\begin{equation}
\Psi=\Phi+\frac{1}{E-H_0}V\Psi \Rightarrow \Psi=\Phi+\frac{1}{E-H_0}T\Phi
\label{eq:wavefun}
\end{equation}
where in the last step of Eq.~\eqref{eq:wavefun} we have used the definition of the scattering matrix $T$, $T\Phi\equiv V\Psi$. 

Eq.~\eqref{eq:wavefun} gives us the full wave function $\Psi$ in terms of free one, $\Phi$, and the scattering matrix. Using the normalization of the state with momentum $\vec p$, used in \cite{Gamermann10}, we have
\begin{equation}
\langle \vec p\,' | \vec p\,\rangle=\delta^3(\vec p - \vec p\,') ; \,\,\, |\vec p\,\rangle \langle \vec p\, | = \int d^3\vec p
\label{eq:norma}
\end{equation}
which implies that in coordinate space
\begin{equation}
\langle \vec r\, |\vec p\,\rangle=\frac{e^{i\vec p\cdot\vec r}}{(2\pi)^{3/2}}
\end{equation}
We are going to use the $T$-matrix obtained as
\begin{equation}
T=\frac{V}{1-VG}
\label{eq:Tmat}
\end{equation}
for a single-channel collision, where $G$ is the loop function of the intermediate particles \cite{Gamermann10}
\begin{equation}
G(E)=\int_{|\vec q\,| < q_\text{max}}\frac{d^3 \vec{q} }{E-\omega_1(q)-\omega_2(q)+i\eta} 
\label{eq:loop}
\end{equation}
with $\omega_i(q)=\sqrt{{\vec q}\,^2+m_i^2}$ the energy of particle $i$. Eq.~\eqref{eq:Tmat} is obtained, as shown in \cite{Gamermann10}, starting from a ultraviolet regularized separable constant (s-wave) potential
\begin{equation}
V(\vec p,\vec p\,')=V\theta(q_\text{max}-|\vec p\,|)\theta(q_\text{max}-|\vec p\,'|)\ ,
\label{eq:seppot}
\end{equation}
which reverts into
\begin{equation}
T(E;\vec p,\vec p\,')=T(E)\theta(q_\text{max}-|\vec p\,|)\theta(q_\text{max}-|\vec p\,'|).
\label{eq:sepT}
\end{equation}
Eq.~\eqref{eq:Tmat}, called on-shell factorization~\cite{Nieves:1999bx} of the Lippmann--Schwinger equation, or Bethe--Salpeter equation if relativistic energies are used, is common in the study of interactions of particles and can be equally justified using dispersion relations and neglecting the contribution of the left hand cut (more properly, including this contribution as an energy independent term \cite{OllerUlf01}). 

Eq.~\eqref{eq:Tmat} is easily generalized to coupled channels as
\begin{equation}
T=\Big[1-VG\Big]^{-1}V
\label{eq:Tmatcoupled}
\end{equation}
where now $V$ is the transition potential $V_{ij}$ between the channels $i$ and $j$ and $G$ is now the diagonal loop function $G\equiv$ diag$[G_i(E)]$, with $G_i(E)$ the loop function of each particular channel $i$. Introducing
the identity operators $|\vec p\,'\rangle\langle \vec p\,'|$ and $|\vec p\,''\rangle\langle \vec p\,''|$ between the terms $(E-H_0)^{-1}T\Phi$ in Eq.~\eqref{eq:wavefun} we obtain
\begin{widetext}
\begin{equation}
\langle \vec r\,|\widetilde{\Psi}\rangle =\frac{e^{i\vec p\cdot\vec r}}{(2\pi)^{3/2}} + 
\theta(q_\text{max}-|\vec p\,|)T(E)\int\frac{d^3\vec q}{(2\pi)^{3/2}}\frac{e^{i\vec q\cdot\vec r}\,\,\theta(q_\text{max}-|\vec q\,|)}{E-\omega_1( q)-\omega_2({ q})+i\eta}\equiv \frac{\langle \vec r\,|\Psi\rangle}{N}\equiv \frac{\Psi(\vec r, \vec p\,)}{N}
\label{eq:wavefun2}
\end{equation}
and the modulus of the relative momenta of the pair of particles is $|\vec{p}\,|=\lambda^{1/2}(s,m^2_1,m^2_2)/(2\sqrt{s})$ being 
$\lambda(a,b,c)=a^2+b^2+c^2-2ab-2ac-2bc$ the well known K\"{a}llen function, and $\sqrt{s}=E$ the total energy in the c.m. frame.  Normalizing the wave function $\Psi(\vec r, \vec p\,)$ to unit flux, as demanded in the construction of Eq.~\eqref{eq:cf}, we fix $N=(2\pi)^{3/2}$ and find 
\begin{equation}
\Psi(\vec r, \vec p\,)=e^{i\vec p\cdot\vec r} + 
\theta(q_\text{max}-|\vec p\,|)\,T(E)\int_{|\vec q\,|< q_\text{max}}\frac{d^3\vec q\ e^{i\vec q\cdot\vec r}}{E-\omega_1( q)-\omega_2({ q})+i\eta}
\label{eq:wavefun3}
\end{equation}
\end{widetext}
%
In the expansion of $e^{i\vec q\cdot\vec r}$ in terms of spherical harmonics, 
\begin{equation}
e^{i\vec q\cdot\vec r}=4\pi\sum_{\ell=0}^{+\infty} i^\ell j_\ell(qr)\sum_{m=-\ell}^{+\ell} Y^*_{\ell m}(\hat q)Y_{\ell m}(\hat r)
\label{eq:spherexp}
\end{equation}
only the $Y^*_{00}(\hat q)$ term contributes in the integral and there, $e^{i\vec q\cdot\vec r}$ is replaced by the spherical Bessel function $j_0(qr)$, 
\begin{equation}
 \Psi(\vec r, \vec p\,)=e^{i\vec p\cdot\vec r} + f_E(r)\
\end{equation}
where the correction to the plane wave incorporates the information on the interaction, and it depends on the energy, the momentum regulator and the relative distance $r$,
\begin{eqnarray}
\hspace{-0.5cm}f_E(r)&=&\theta(q_\text{max}-|\vec p\,|)\,T(E)\times\nonumber \\
&\times& \int_{|\vec q\,|< q_\text{max}}\frac{ d^3\vec q\,j_0(qr)}{E-\omega_1( q)-\omega_2({ q})+i\eta} 
\label{eq:fr}
\end{eqnarray} 
Thus, the modulus square of the wave function can be written simply as
\begin{equation}  
|\Psi(\vec r,\vec p\,)|^2=1+|f_E(r)|^2+2\mbox{Re}\Big[e^{i\vec p\cdot\vec r}f^*_E(r)\Big] \ .
\label{eq:fr2}
\end{equation}

When performing $\int d^3\vec r S_{12}(r)2\mbox{Re}\Big[e^{i\vec p\cdot\vec r}f^*_E(r)\Big]$, we can again expand $e^{i\vec p\cdot\vec r}$ in terms of spherical harmonics of $\vec r$ as done in Eq.~\eqref{eq:spherexp} and since $S_{12}(r)$ does not depend on angular variables, we again pick up the $Y_{00}(\hat r)$ term and therefore
\begin{align}
\hspace{-0.25cm}&\int d^3\vec r\, S_{12}(r)|\Psi(\vec r,\vec p\,)|^2 = 1+4\pi\int_0^{+\infty} drr^2\,S_{12}(r) \nonumber \\
&\hspace{1cm}\times \Big\{|j_0(pr)+f_E(r)|^2-j^2_0(pr) \Big\} \label{eq:cf2}
\end{align}
obtaining the Koonin--Pratt formula \cite{Koonin77,Pratt90,Bauer92}. Note that in the derivation of Eq.~\eqref{eq:cf2} we added and subtracted $j^2_0(pr)$.

With respect to the Koonin--Pratt formula we obtain (see Eq.~\eqref{eq:fr}) the $d^3\vec q$ integral regulated by the maximum momentum $q_\text{max}$ demanded by our approach for the $T$-matrix and the factor $\theta(q_\text{max}-|\vec p\,|)$ which will be inoperative in the range of values of 
the relative momentum $p$ of interest for the correlation function, with values of $p$ smaller than $q_\text{max}$. Note that $f_E(r)$ in Eq.~\eqref{eq:fr} is already regulated by the Bessel function $j_0(qr)$ and we shall investigate this in connection with the range of the interaction introduced by our formalism.

In the next step we introduce the effect of coupled channels, similarly as done in Ref.~\cite{Haidenbauer19}. Let us assume we have $n$ coupled channels and that the channel $i$ is the one observed in the final state. We can reach this state starting from any other channel $j\neq i$, and by analogy to Eq.~\eqref{eq:wavefun3} the corresponding wave function will be
\begin{align}
\hspace{-0.25cm}&\Psi_j(\vec r, \vec p\,)=T_{ji}(E)\theta(q_\text{max}-|\vec p\,|)\nonumber\\
&\times\int_{|\vec q\,|< q_\text{max}} \,\frac{d^3\vec q\, e^{i\vec q\cdot\vec r}}{E-\omega^{(j)}_1(q)-\omega^{(j)}_2({q})+i\eta}
\end{align} 
where $\omega^{(j)}_1(q)$ and $\omega^{(j)}_2(q)$ are the kinetic energies,including masses, of particles $1,2$ in the channel $j$ and $E$ and $p$ are related through $E = \omega^{(i)}_1(p) + \omega^{(i)}_2(p)$. Note that in the $d^3\vec q$ integral, one can again replace  $e^{i\vec q\cdot\vec r}$ by $j_0(qr)$.
In the Koonin--Pratt formalism the contribution from these channels is added incoherently to the one of the observed channel, with a weight $w_j$ relative to the weight $w_i=1$ implicit in Eq.~\eqref{eq:cf2} for the observed channel. We can write in general for the $i$ observed channel 
\begin{align}
\hspace{-0.25cm}&\Psi_j(\vec r, \vec p\,)=\delta_{ij}j_0(pr)+T_{ji}(E)\theta(q_\text{max}-|\vec p\,|)\nonumber \\
&\times  \int_{|\vec q\,|< q_\text{max}} d^3\vec q \frac{j_0(qr)}{E-\omega_1^{(j)}(q)-\omega_2^{(j)}(q)+i\eta}
\label{eq:wavefun5}
\end{align}  
and the correlation function will read
\begin{align}
\hspace{-0.25cm}&C(\vec p\,)=1+4\pi\int_0^{+\infty} drr^2 S_{12}(r)\nonumber \\
&\times\left(\sum_jw_j|\widetilde{\Psi}_j(\vec r,\vec p\,)|^2-j^2_0(pr) \right) 
\label{eq:cf3}
\end{align}   
with the weight $w_i$ of the observed channel $i$ equal to $1$. This formula is also used in Ref.~\cite{Wei23} to evaluate the correlation function in the case of the $DK$ interaction with a different normalization which we address below.

In the $T_{cc}$ study of Ref.~\cite{Feijoo21} a different, relativistic normalization is used for the $T$-matrix, common in quantum field theory approaches. Since $VG$ (or $TG$) is weighted versus $1$ in Eq.~\eqref{eq:Tmatcoupled}, the product $TG$ is independent of normalization conventions. Since the two-hadron loop function in \cite{Feijoo21} is given by \cite{Oller97}
\begin{align}
&G^\text{FT}(\sqrt{s}=E)=\int \frac{d^3\vec q}{(2\pi)^3}\frac{\omega_1(q)+\omega_2(q)}{2\omega_1(q)\,\omega_2(q)}\nonumber \\
&\times\frac{1}{s-\left[\omega_1(q)+\omega_2({ q})\right]^2+i\eta} \,\,,
\label{eq:GFT}
\end{align} 
comparing this loop function with the one so far used of Eq.~\eqref{eq:loop} we can immediately write with the new conventions\footnote{In Eq.~\eqref{eq:wavefun6}, we should have used the notation $T^\text{FT}$, referring to the scattering amplitude obtained in Ref.~\cite{Feijoo21}, but for simplicity, the FT tag is not included. In addition,  we will also omit, from now on, the FT label in the loop function $G$. The relation of $T^\text{FT}$ with the $S-$matrix can be found \textit{e.g.} in Ref.~\cite{Gamermann10} (see also next Sec.~\ref{sec:tcc}).}
\begin{align}
&\Psi_j(\vec r, \vec p\,)=\delta_{ij}j_0(p r) + T_{ji}(E)\theta(q_\text{max}-|\vec p\,|)\times\nonumber \\
&\int_{|\vec q\,|< q_\text{max}} \frac{d^3\vec q}{(2\pi)^3}\frac{\omega_1^{(j)}( q)+\omega_2^{(j)}( q)}{2\omega_1^{(j)}( q)\,\omega_2^{(j)}( q)} \nonumber \\
&\times \frac{j_0(qr)}{s-\left[\omega_1^{(j)}(q)-\omega_2^{(j)}(q)\right]^2+i\eta}~,
\label{eq:wavefun6}
\end{align}  
and Eq.~\eqref{eq:cf3} equally holds with the new conventions. Under the approximation $s-(\omega_1+\omega_2)^2\sim 2(\omega_1+\omega_2)(\sqrt{s}-\omega_1-\omega_2)$ one recovers the formulas used in Ref.~\cite{Wei23}.
  
\section{The $T_{cc}(3875)^+$ states with $D^0D^{*+}$ and $D ^+D^{*0}$ channels }
\label{sec:tcc}

In Ref.\ \cite{Feijoo21} the interaction between the $D^0D^{*+}$ and $D ^+D^{*0}$ channels was studied exchanging vector mesons in the extension of the local hidden gauge symmetry approach \cite{Bando88,Harada03,Meissner88,Nagahiro09} to the charmed sector. 
It was found that the isospin $I=0$ combination had an attractive interaction which would bind the system, while the $I=1$ one had a repulsive interaction. This observation agrees with the conclusions of the experimental analysis \cite{Aaij22c}.  The study of Ref.\ \cite{Feijoo21} was done with the explicit two channels where the $T_{ij}(\sqrt{s}=E)$ amplitudes were evaluated. Only fine tuning of the parameter $q_\text{max}$ was done to obtain the experimental binding of \cite{Aaij22c}, and the width came out automatically from the consideration of the $D^*\rightarrow D\pi$ decay. We take advantage to mention, since this information will be used below, that from this work we can find the scattering lengths and effective ranges
\begin{eqnarray}
a_{D^0D^{*+}}&=&(7.78-1.82\,i) \,\,\mbox{fm} \nonumber \\
a_{D^+D^{*0}}&=&(2.07-1.21\,i) \,\,\mbox{fm} \nonumber \\
r_{0,D^0D^{*+}}&=&-3.49 \,\,\mbox{fm} \nonumber \\
r_{0,D^+D^{*0}}&=&(-0.06-2.27\,i) \,\,\mbox{fm} 
\label{eq:scattlen}
\end{eqnarray}
where in the evaluation of $r_0$ the very small $D^*$ width is neglected,\footnote{This is sufficiently accurate for the purpose of this work. Nevertheless, a complete study of the three-body dynamics in the $T_{cc}$ due to the finite life time of the $D^*$ can be found in Ref.~\cite{Du22}. There, the pion exchange between the $D$ and $D^*$ mesons and the finite $D^*$ width, are taken into account simultaneously to ensure that three-body unitarity is preserved.  In that work the low-energy expansion of the $DD^*$ scattering amplitude is also performed and the scattering length and effective range are extracted maintaining finite the width of the $D^*$.} and
where $a$ and $r_0$ are defined from the Quantum Mechanics amplitude in the effective range expansion around threshold $\sqrt{s_\text{th}}$, where $p$ vanishes
\begin{equation} 
f_\text{QM} \simeq \frac{1}{-\frac{1}{a}+\frac{1}{2}r_0p^2-ip} \ .
\label{eq:famplitude}
\end{equation}
It is easy to relate our scattering matrix with this one since from Eqs.~\eqref{eq:Tmatcoupled}  and \eqref{eq:GFT} with a real potential one has
%
 \begin{equation}
 \mbox{Im}\,T^{-1} = -\mbox{Im}\,G = \frac{p}{8\pi\sqrt{s}}
\label{eq:ImagT}
\end{equation}
On the other hand
\begin{equation}
\mbox{Im}\,f^{-1}_\text{QM}=-p \ .
\label{eq:Imagf}
\end{equation}
Hence
\begin{equation}
T = -8\pi\sqrt{s}f_\text{QM} \Rightarrow a=\frac{T(\sqrt{s_\text{th}})}{8\pi\sqrt{s}} \ .
\label{eq:rel_Tf}
\end{equation} 
The values reported in Eq.~\eqref{eq:scattlen} for the scattering length agree well with those extracted from experiment in \cite{Aaij22c}
\begin{eqnarray}
a^\text{exp}_{D^0D^{*+}}&=&(7.16\pm 0.51-1.85\pm 0.28\,i) \,\,\mbox{fm} \nonumber \\
a^\text{exp}_{D^+D^{*0}}&=&(1.76 -1.82 \,i) \,\,\mbox{fm} \nonumber \ , \\
\label{eq:scattlen_exp}
\end{eqnarray} 
with similar relative errors expected for $a^\text{exp}_{D^+D^{*0}}$. These values also agree with those obtained in Refs.~\cite{Albaladejo22,Du22}.

In Ref.~\cite{Kamiya22} a scheme is used in which the potentials are written in coordinate space and the formalism is conducted using explicitly wave functions as solutions of the Sch\"{o}dinger equation. The range of the potential is assumed to be given by one pion exchange, which is drastically different from the one used in \cite{Feijoo21}, where the interaction originates from $\rho$ exchange and the range in momentum space is given by $q_{\rm max}\sim 420$ MeV fitted to data. We shall see that our results for the correlation function are very similar, which might be surprising in view of the different ranges assumed for the interaction. We shall find an explanation for this surprising result in the fact that the strength of the potentials in \cite{Kamiya22} is taken\footnote{ Note that in the approach of Ref.~\cite{Feijoo21}, the $DD^*$ contact interaction is fixed to the prediction of the hidden gauge model.} such as to fit the experimental scattering lengths reported in Eq.~\eqref{eq:scattlen_exp}, and the correlation functions in the present case are practically only sensitive to the scattering length.

To finish this section let us write explicitly the correlation function for the $D^0 D^{*+}$ and $D ^+ D^{*0}$ coupled-channel problem that we are discussing,
\begin{align}
\hspace{-0.25cm}&C_{D^0D^{*+}}(p_{D^0})=1+4\pi\,\theta(q_\text{max}-p_{D^0}) \times\nonumber \\
&\int_0^{+\infty} drr^2S_{12}(r)\Big\{\big|j_0(p_{D^0}r)+T_{11}(E)\widetilde{G}^{(1)}(r;E)\big|^2 \nonumber \\
&+\big|T_{21}(E)\widetilde{G}^{(2)}(r;E)\big|^2-j^2_0(p_{D^0}r)\Big\}
\label{eq:cf4}
\end{align} 
for detecting the $D^0D^{*+}$ pair, with $E=\sqrt{s}=\left[\sqrt{m_{D^0}^2+p_{D^0}^2}+\sqrt{m_{D^{*+}}^2+p_{D^{0}}^2})\right]$,
and
\begin{align}
\hspace{-0.25cm}&C_{D^+D^{*0}}(p_{D^+})=1+4\pi\,\theta(q_\text{max}-p_{D^+}) \times\nonumber \\
&\int_0^{+\infty} drr^2S_{12}(r)\Big\{\big|j_0(p_{D^+}r)+T_{22}(E)\widetilde{G}^{(2)}(r;E)\big|^2 \nonumber \\
&+\big|T_{12}(E)\widetilde{G}^{(1)}(r;E)\big|^2-j^2_0(p_{D^+}r)\Big\}
\label{eq:cf5}
\end{align} 
for observing $D^+D^{*0}$, where now the c.m. energy is given by  $E=\left[\sqrt{m_{D^+}^2+p_{D^+}^2}+\sqrt{m_{D^{*0}}^2+p_{D^{+}}^2})\right]$.

Finally, the quantities $\widetilde{G}^{(i)}$ are simply
\begin{align}
\widetilde{G}^{(i)}(r;E)&=\int_{|\vec q\,|< q_\text{max}} \frac{d^3\vec q}{(2\pi)^3}\frac{\omega^{(i)}_1(q)+\omega^{(i)}_2(q)}{2\omega^{(i)}_1(q)\,\omega^{i}_2(q)}\nonumber \\
&\times\frac{j_0(qr)}{s-\left[\omega_1^{(i)}(q)+\omega_2^{(i)}({q})\right]^2+i\eta}
\label{eq:Gtilde}
\end{align} 
and the $T$-matrix elements, $T_{ij}(E)$, are taken from \cite{Feijoo21}. We have taken $(w_1=1)$ for the observed channel, as we discussed, and for the other channel, given the analogy between $D^0D^{*+}$ and $D ^+D^{*0}$, we have also taken the weight equal to one, $w_2=1$.

\section{Results}
\label{sec:results}

\begin{center}
\begin{figure}[t]
\begin{center}
\includegraphics[width=0.4\textwidth,keepaspectratio]{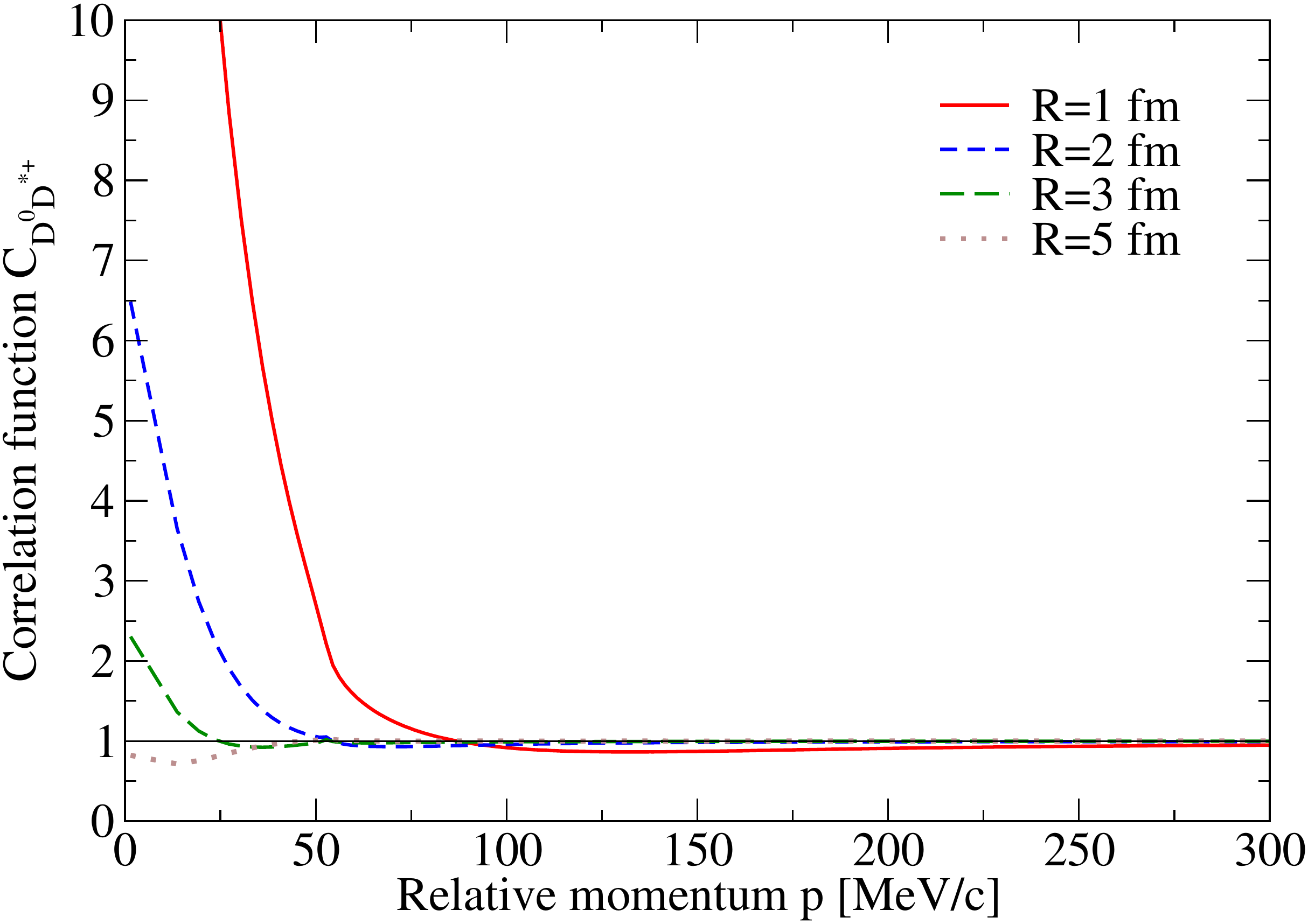}
\caption{(Color online). Correlation function of the $D^0D^{*+}$ pair for different values of the source size.}
\label{fig:fig1}
\end{center}
\end{figure}
\end{center}
\begin{center}
\begin{figure}[t]
\begin{center}
\includegraphics[width=0.4\textwidth,keepaspectratio]{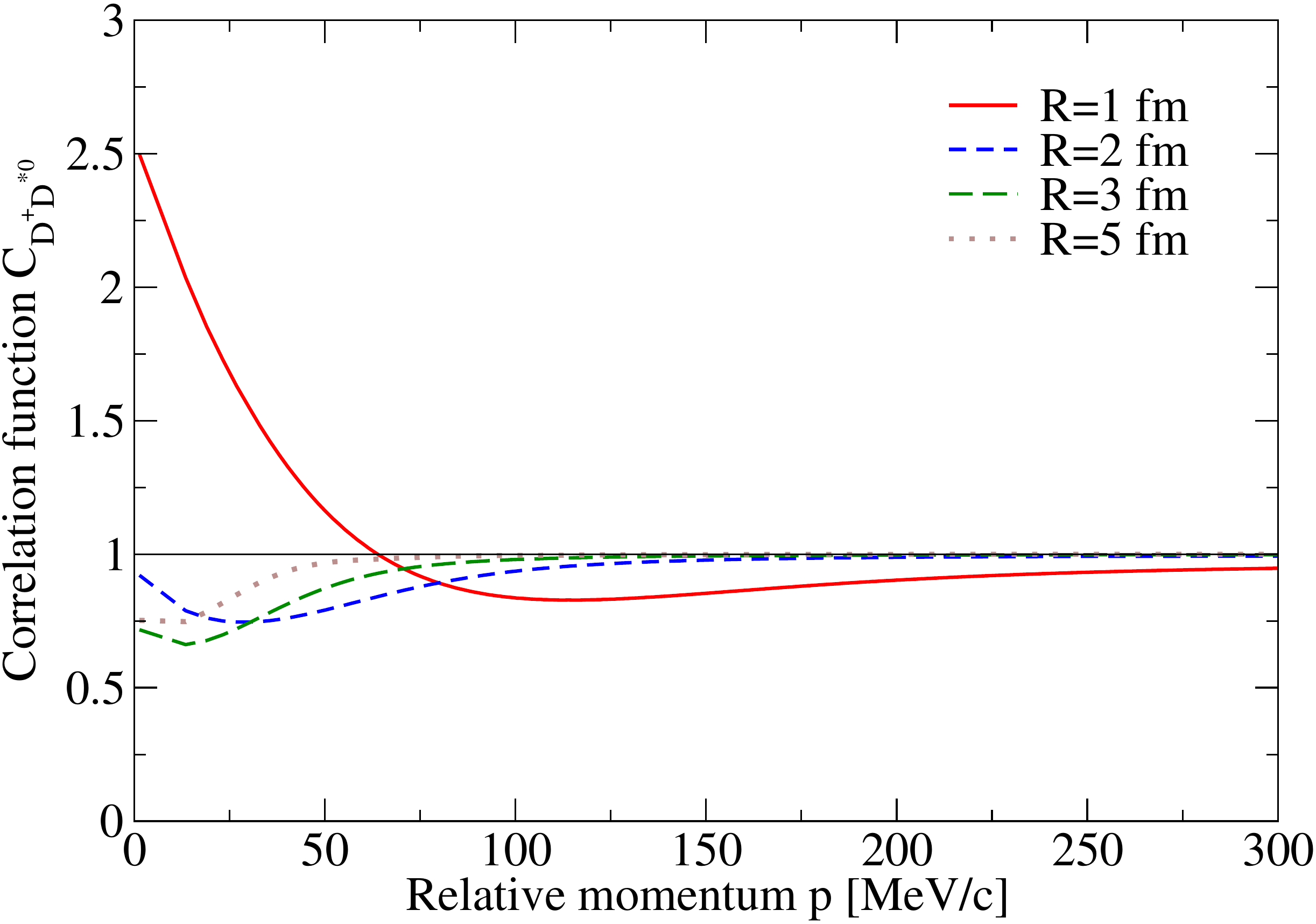}
\caption{(Color online). Correlation function of the $D^+D^{*0}$ pair for different values of the source size.}
\label{fig:fig2}
\end{center}
\end{figure}
\end{center}
In Figs.\ \ref{fig:fig1} and \ref{fig:fig2} we show our results for the correlation function for the $D^0D^{*+}$ and $D ^+D^{*0}$ channels
respectively, and different values of the source size $R=1,2,3,5$ fm. We can see that for the case $D^0D^{*+}$, the correlation function at very small $p$ and $R=1$ fm is quite large compared to 1. This is also what is found in Ref.~\cite{Kamiya22}. As the source size $R$ is increased, the correlation function becomes closer to unity at smaller momenta. We also observe a fast convergence of the correlation function to unity for values of $p$ around 200 MeV. As anticipated, the factor $\theta(q_\text{max}-|\vec p\,|)$ in Eq.~\eqref{eq:wavefun6} is inoperative, since $q_\text{max}=420$ MeV in \cite{Feijoo21} and we only go in both figures up to 300 MeV. Also in the case of the $D ^+D^{*0}$, the correlation function obtained is comparable with the one calculated in \cite{Kamiya22}, and features similar to the ones just described for the correlation function of the $D^0D^{*+}$ pair are also observed in this case. As we shall see below, the $T_{cc}$ correlation functions are almost determined by the scattering lengths and unitarity. Thus, although the interaction range in coordinate space of \cite{Kamiya22} is much larger than that in the approach of  Ref.~\cite{Feijoo21}, the $D D^*$ isoscalar interaction strength (the sharp ultraviolet cutoff $q_\text{max}$) is adjusted in the former (latter) work to reproduce the experimental $T_{cc}$ mass. Thus both approaches, the one of Ref.~\cite{Kamiya22} and that of Ref.~\cite{Feijoo21}, provide similar scattering lengths and hence correlation functions. In Sec.~\ref{sec:compar} we will also compute the correlation functions using the $T$-matrix obtained in Ref.~\cite{Albaladejo22}, finding also similar results due to the same argument.

\begin{center}
\begin{figure}[t]
\begin{center}
\includegraphics[width=0.4\textwidth,keepaspectratio]{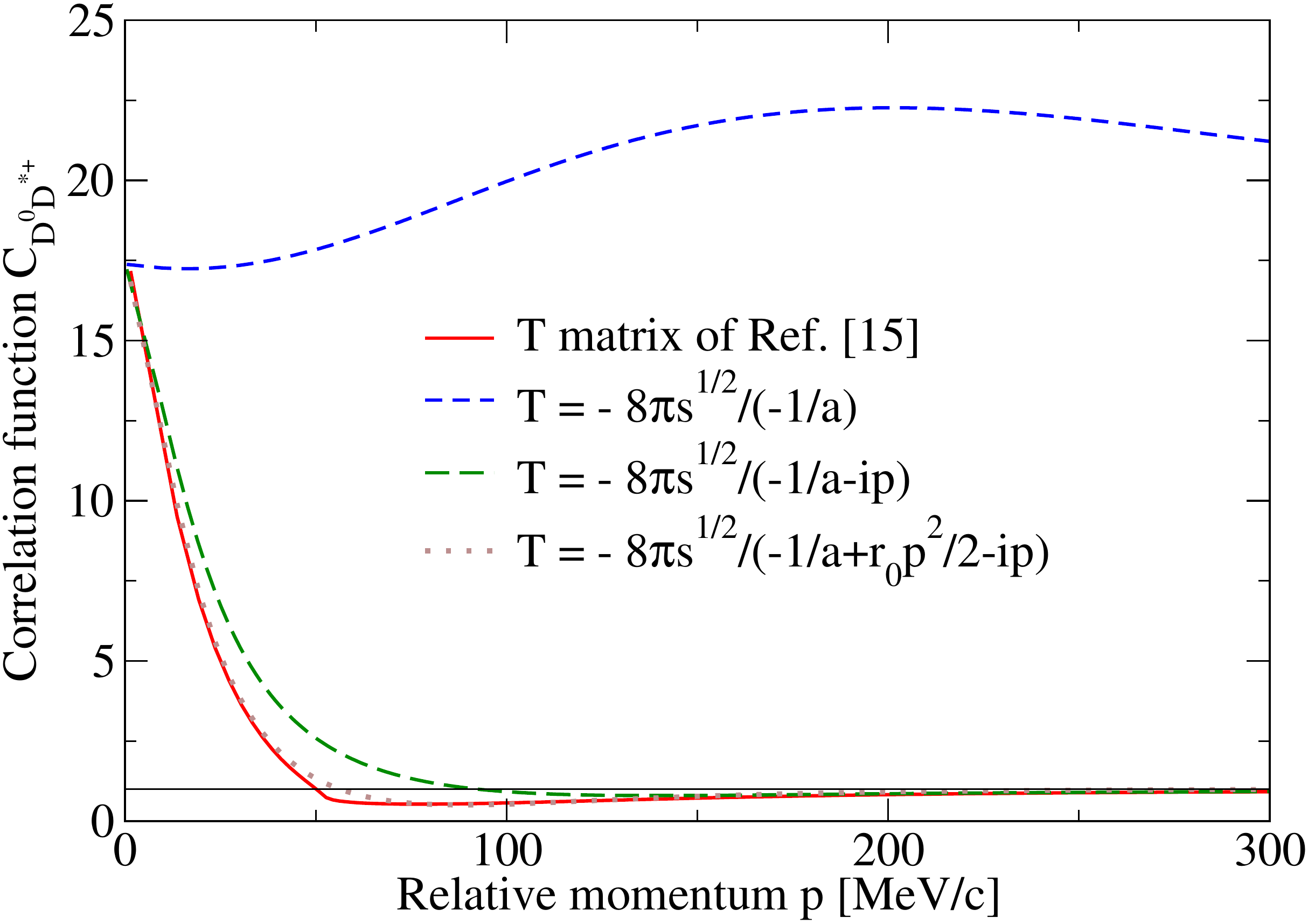}
\caption{(Color online). Correlation function of the $D^0D^{*+}$ pair obtained with the $T$-matrix of Ref.\ \cite{Feijoo21} and different approximations in terms of the scattering length and the effective range. The source size is taken $R=1$ fm.}
\label{fig:fig3}
\end{center}
\end{figure}
\end{center}
\begin{center}
\begin{figure}[t]
\begin{center}
\includegraphics[width=0.4\textwidth,keepaspectratio]{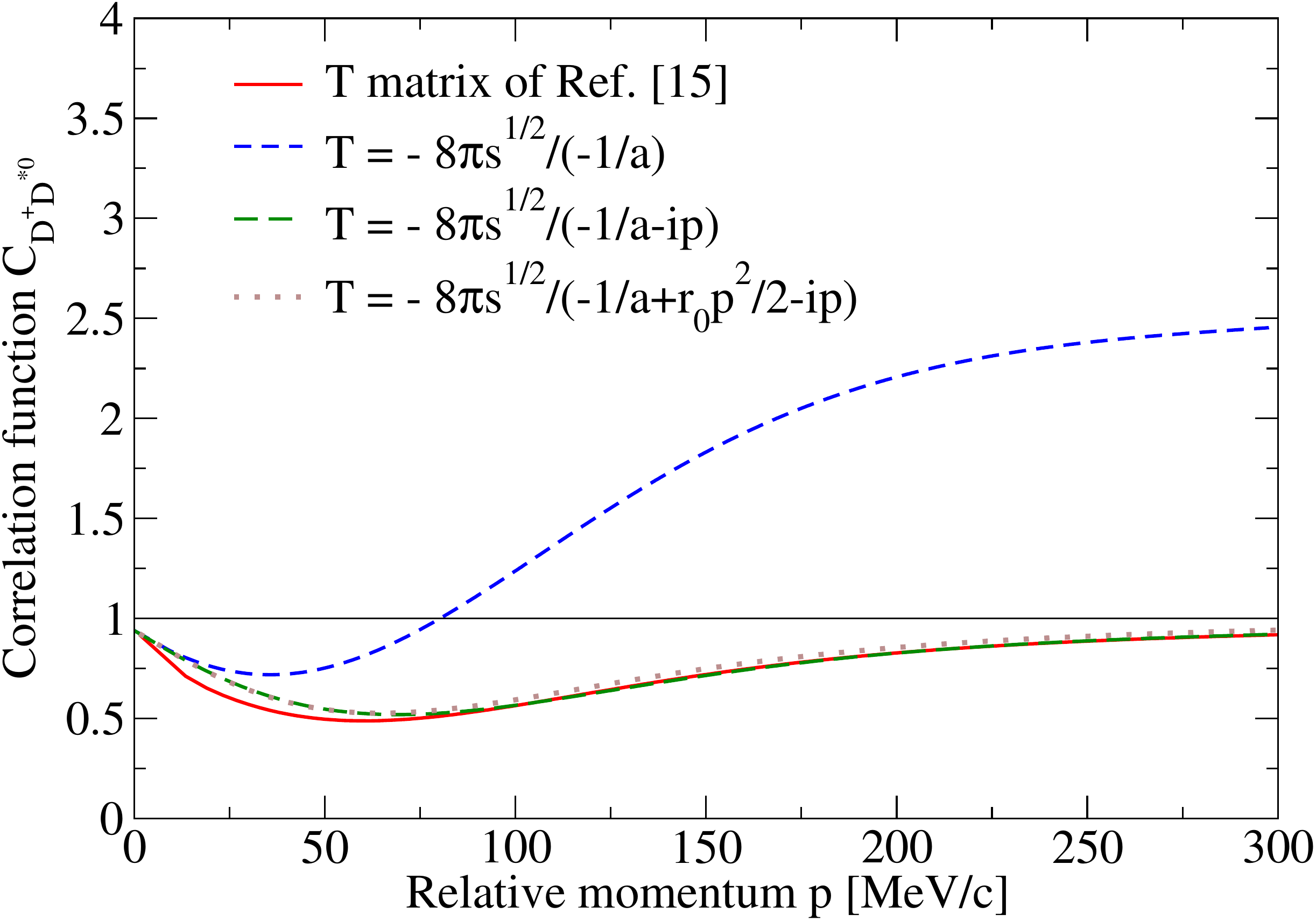}
\caption{(Color online). Correlation function of the $D^+D^{*0}$ pair obtained with the $T$-matrix of Ref.\ \cite{Feijoo21} and different levels of approximation in terms of the scattering length and the effective range. The source size is taken $R=1$ fm.}
\label{fig:fig4}
\end{center}
\end{figure}
\end{center}

The next step is to see to which magnitudes of the scattering matrix the correlation function is sensitive. For this purpose we take $R=1$ fm and show in Figs.~\ref{fig:fig3} and \ref{fig:fig4} the results using $T=-8\pi\sqrt{s}/(-1/a)$  (blue dashed lines), $T=-8\pi\sqrt{s}/(-1/a-ip)$ (green long dashed lines) and $T=-8\pi\sqrt{s}/(-1/a+r_0p^2/2-ip)$ (brown dotted lines) and compare the results with those obtained using the full $T$-matrix evaluated in \cite{Feijoo21}. Note that since the scattering length and the effective range are defined only for the diagonal channels $D^0D^{*+}\rightarrow D^0D^{*+}$ and $D ^+D^{*0}\rightarrow D ^+D^{*0}$, in Figs.~\ref{fig:fig3} and \ref{fig:fig4} for the calculation of the correlation function with the $T$-matrix of \cite{Feijoo21} we have taken, respectively, the weights of the wave functions $\Psi_{D^+D^{*0}\rightarrow D^0D^{*+}}$ and $\Psi_{D^0D^{*+}\rightarrow D^+D^{*0}}$ equal to zero ($w_2=0$). This is to say, $T_{21}$ and $T_{12}$ effects are neglected in Eqs.~\eqref{eq:cf4} and \eqref{eq:cf5}, respectively. We can see that for the approximation $T \propto a$ we only get the result of the correlation function at threshold, as expected, but as $p$
increases we see that the correlation function does not converge to 1 as it should. However, when we use $T \propto (-1/a-ip)^{-1}$
the agreement with the exact calculation is very good, with differences which are smaller than ordinary experimental errors, which indicates that the correlation function in both cases gives us information mostly about the scattering length. Yet, for the correlation function of the $D^0D^{*+}$ channel, the consideration of the effective range $r_0$ helps to improve the agreement with the exact result. The message then is that, provided one finds experimental input to choose the right parameter of the source, $R$, we could then get the scattering length for the $D^0D^{*+}$ and $D ^+D^{*0}$ channels, but probably less information about higher order parameters in the effective range expansion. Determining the scattering length, of course, is a very valuable information, and it seems a rather general feeling about what can be accomplished with correlation functions, although for particular cases one may get a different  conclusion.

\begin{center}
\begin{figure}[t]
\begin{center}
\includegraphics[width=0.4\textwidth,keepaspectratio]{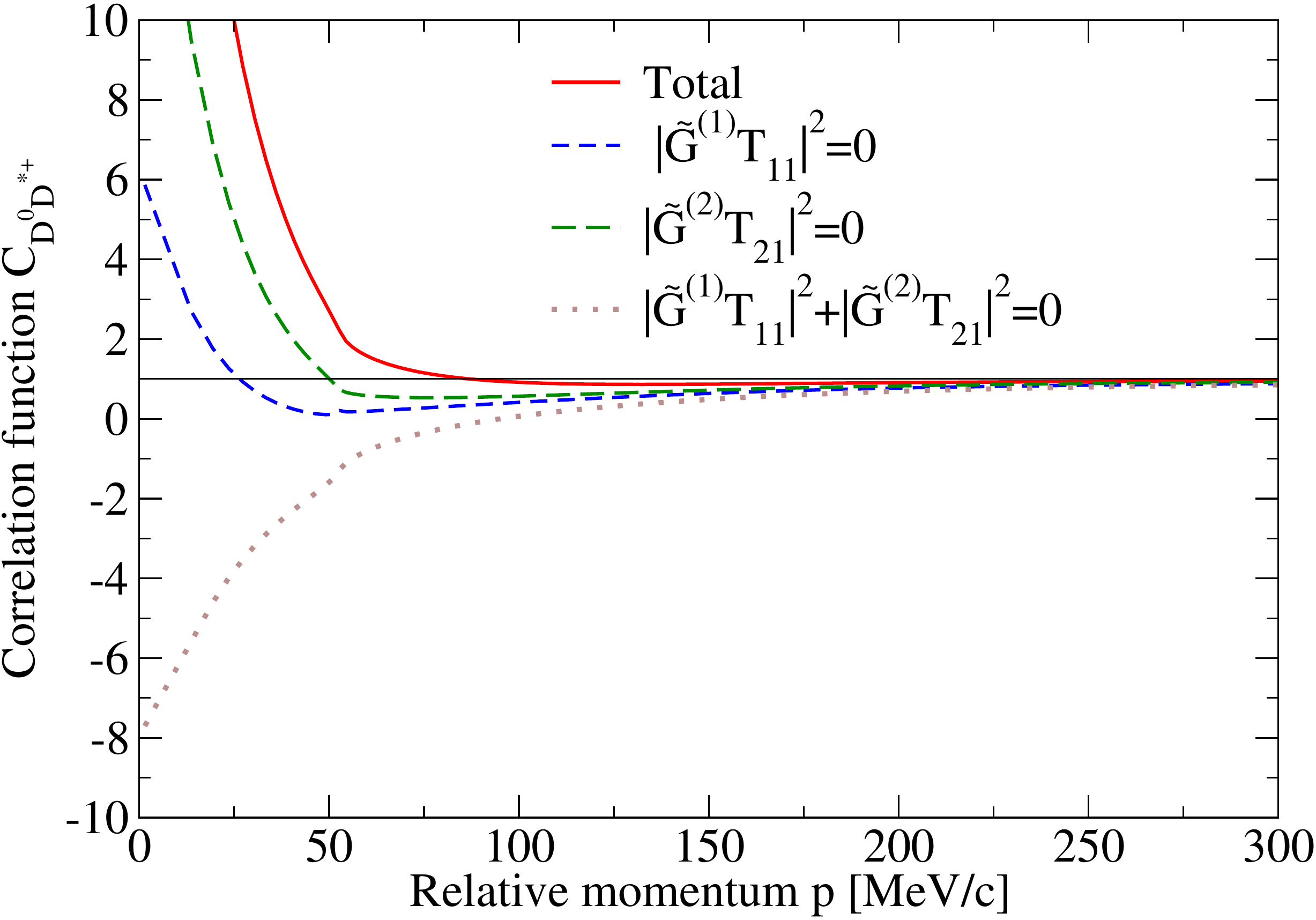}
\caption{(Color online). Different contributions to the correlation function of the pair $D^0D^{*+}$. The source size is taken $R=1$ fm.}
\label{fig:fig5}
\end{center}
\end{figure}
\end{center}

\begin{center}
\begin{figure}[t]
\begin{center}
\includegraphics[width=0.4\textwidth,keepaspectratio]{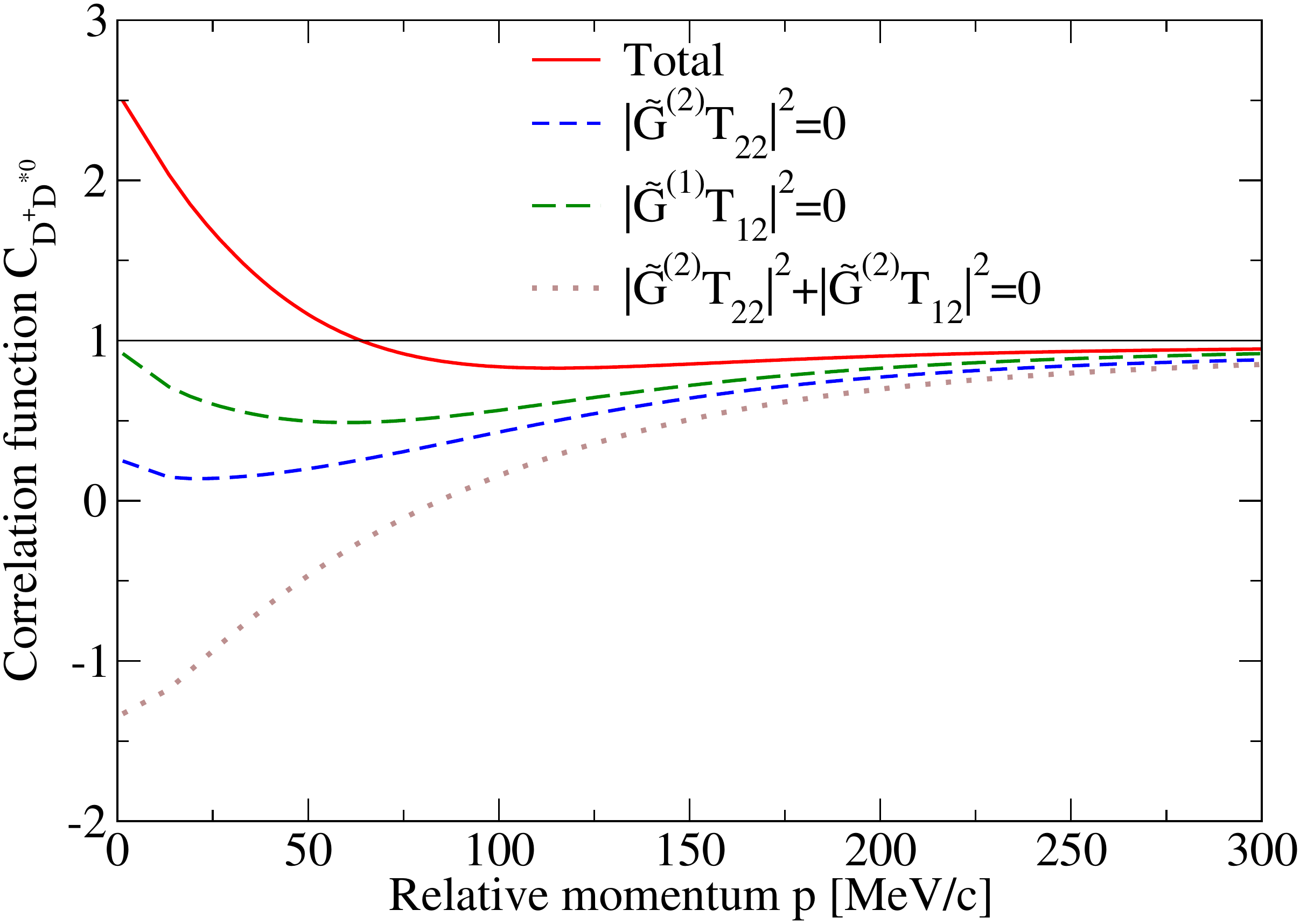}
\caption{(Color online). Different contributions to the correlation function of the pair $D^+D^{*0}$. The source size is taken $R=1$ fm.}
\label{fig:fig6}
\end{center}
\end{figure}
\end{center}

It is interesting to see where the contribution to the correlation function comes from. For this purpose in Figs.~\ref{fig:fig5} and \ref{fig:fig6} we show the results removing in Eqs.~\eqref{eq:cf4} and \eqref{eq:cf5}, respectively, the first channel $|T_{11}\widetilde{G}^{(1)}|^2$ and $|T_{22}\widetilde{G}^{(2)}|^2$ (blue dashed lines), the second channel, $|T_{21}\widetilde{G}^{(2)}|^2$ and $|T_{12}\widetilde{G}^{(1)}|^2$  (green long dashed lines), and also keeping only the interference term between the incoming and the outgoing wave of the observed channel (dotted lines). This is subtracting $|T_{11}\widetilde{G}^{(1)}|^2$ and $|T_{21}\widetilde{G}^{(2)}|^2$ from Eq.~\eqref{eq:cf4} and
$|T_{22}\widetilde{G}^{(2)}|^2$ and $|T_{12}\widetilde{G}^{(1)}|^2$ in Eq.~\eqref{eq:cf5}. We observe that taking into account the re-scattering term in the observed channel is essential to go from negative $C(p)$ to positive at small values of $p$ in the case of $D^0D^{*+}$. The contribution of the re-scattering of the non observed channel is also important but less than the rescattering of the observed channel. 


\begin{center}
\begin{figure}[t]
\begin{center}
\includegraphics[width=0.4\textwidth,keepaspectratio]{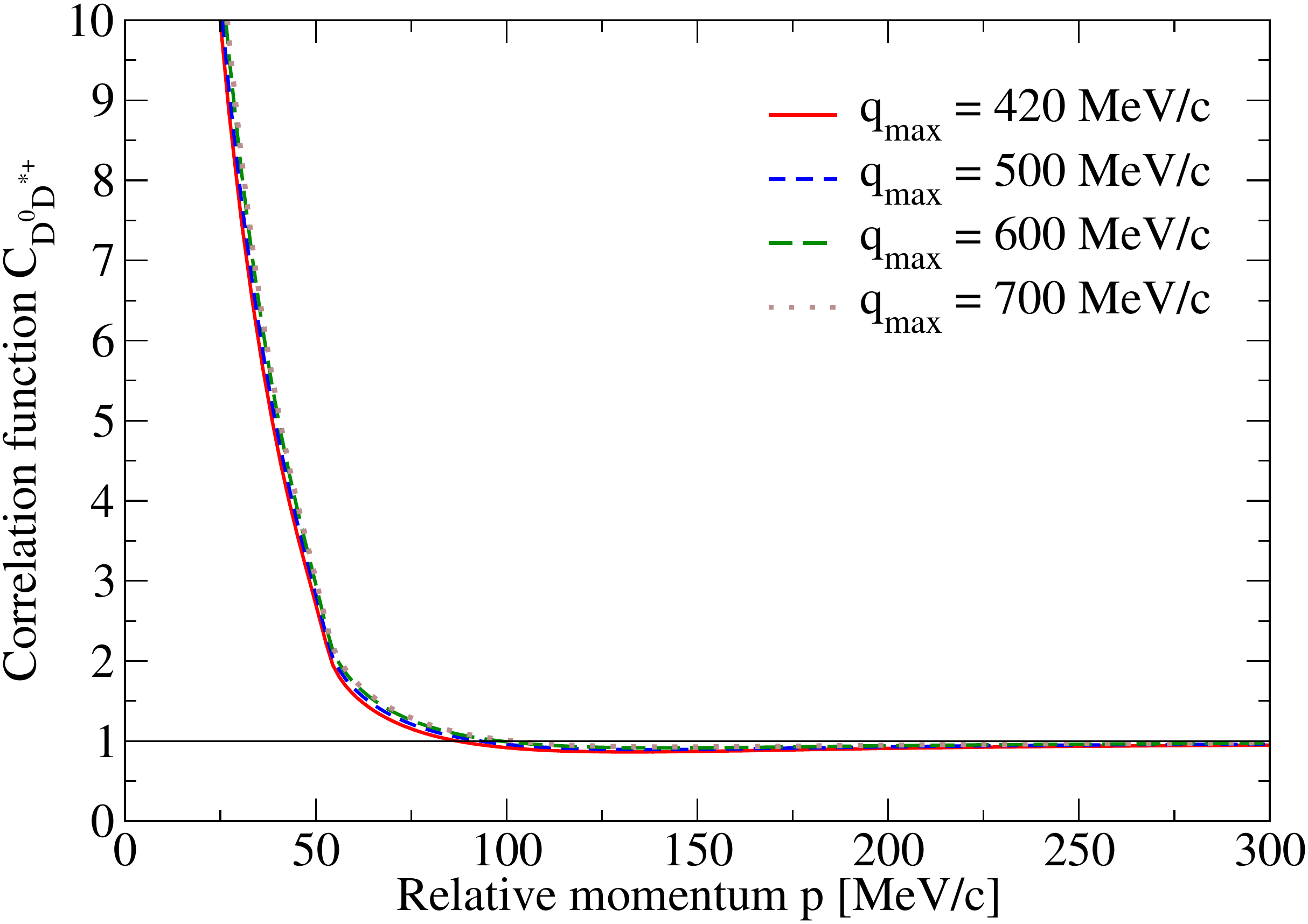}
\caption{(Color online). Dependence of the correlation function of the pair $D^0D^{*+}$ on the cutoff parameter $q_\text{max}$. The source size is taken $R=1$ fm.}
\label{fig:fig7}
\end{center}
\end{figure}
\end{center}

\begin{center}
\begin{figure}[t]
\begin{center}
\includegraphics[width=0.4\textwidth,keepaspectratio]{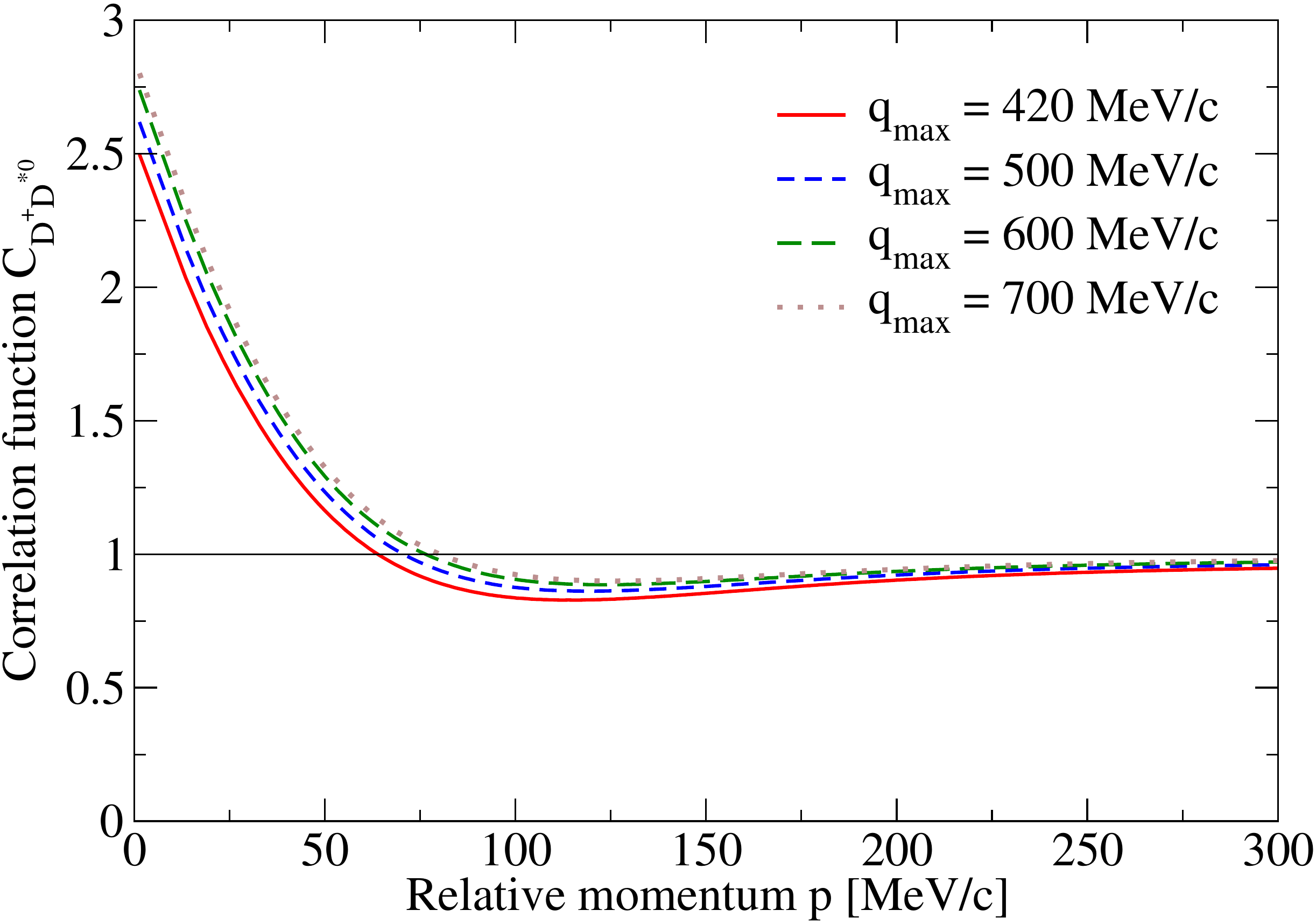}
\caption{(Color online). Dependence of the correlation function of the pair $D^+D^{*0}$ on the cutoff parameter $q_\text{max}$. The source size is taken $R=1$ fm.}
\label{fig:fig8}
\end{center}
\end{figure}
\end{center}

Finally, we present here another test. One novelty of our approach is that we regularize $\widetilde{G}$ with the same cutoff that is used to regularize $G$ when calculating the scattering in \cite{Feijoo21}. Although for consistency we have to use the same cutoff in $G$ and $\widetilde{G}$ as we have shown in the derivation, it is instructive to see how sensitive is the correlation function to the parameter $q_\text{max}$, which as shown in Eq.~\eqref{eq:seppot} provides the range of the interaction in momentum space. We show in Figs.~\ref{fig:fig7} and \ref{fig:fig8} what happens if we increase this cutoff in the evaluation of $\widetilde{G}$ in Eq.~\eqref{eq:Gtilde}, while keeping the $T_{ij}$ matrices of \cite{Feijoo21} unchanged. What we see is that the results are rather insensitive to the value of the cutoff, particularly in the case of the correlation function of the $D^0D^{*+}$ pair, which again shows the limitations of the correlation function to provide information on this magnitude. Note that we are changing $q_\text{max}$ only in $\widetilde{G}$, but $q_\text{max}$ enters the evaluation of the scattering matrix. This has also a positive reading, since if one wishes to obtain the $T_{ij}$-matrix from experimental correlation functions, the range of the interaction needed in the evaluation of $\widetilde{G}$ does not matter much. 


It is interesting to see the reason for this relative lack of sensitivity. Let us take for instance the term $2j_0(p_{D^0}r)\mbox{Re}[T_{11}\widetilde{G}^{(1)}]$ in Eq.~\eqref{eq:cf4}. We have
\begin{equation}
\int d^3\vec q\int drr^24\pi S_{12}(r)j_0(p_{D^0}r)j_0(qr)\,\cdot\cdot\cdot\cdot 
\label{eq:termj0reGT}
\end{equation}
We can write 
\begin{equation}
j_0(qr) = \frac{e^{iqr}-e^{-iqr}}{2iqr}
\label{eq:Bessj0}
\end{equation}
and the same for $j_0(p_{D^0}r)$ and then perform the integration over $r$. We obtain some polynomial and most importantly the exponential factor
\begin{equation}
e^{-Q^2R^2}
\label{eq:exponential}
\end{equation}
with $Q=p_{D^0}+q$ and $Q=p_{D^0}-q$. With a value of $R=1$ fm this factor kills values of $q$ of the order of $R^{-1}\approx 200$ MeV in the integration over $q$. The Bessel function acts as a regulator of the $d^3\vec q$ integral and if we cut it with a value of $q_\text{max}$ reasonably larger than $200$ MeV this extra regulator becomes inoperative. In other words, the correlation function will provide little information on the range of the interaction in momentum space, unless this is reasonably smaller than $R^{-1}$. This also means that with values of $R$ of the order of $5$ fm any possible information on the range of the interaction is greatly lost. The argumentation would proceed similarly with the quadratic terms in the scattering matrix, where we would have $\int d^3\vec q\int d^3q'$ and the same factor of Eq.~\eqref{eq:exponential} with $Q=q+q', Q=q-q'$. Certainly, the correlation functions can provide information on the $T_{ij}$ matrix which depends on the range of the interaction through $q_\text{max}$ in $G$ of Eq.~\eqref{eq:Tmatcoupled}, but since we can get similar $T$-matrices by a simultaneous change of the strength of the potential and $q_\text{max}$, such that $\delta(VG)=0$, once again, it is not easy to obtain information on this interaction range, as illustrated above in Figs.~\ref{fig:fig3} and \ref{fig:fig4}.

\vspace{0.5cm}
\section{Comparison with other schemes}
\label{sec:compar}

\begin{figure*}
    \centering
    \includegraphics[width=0.4\textwidth]{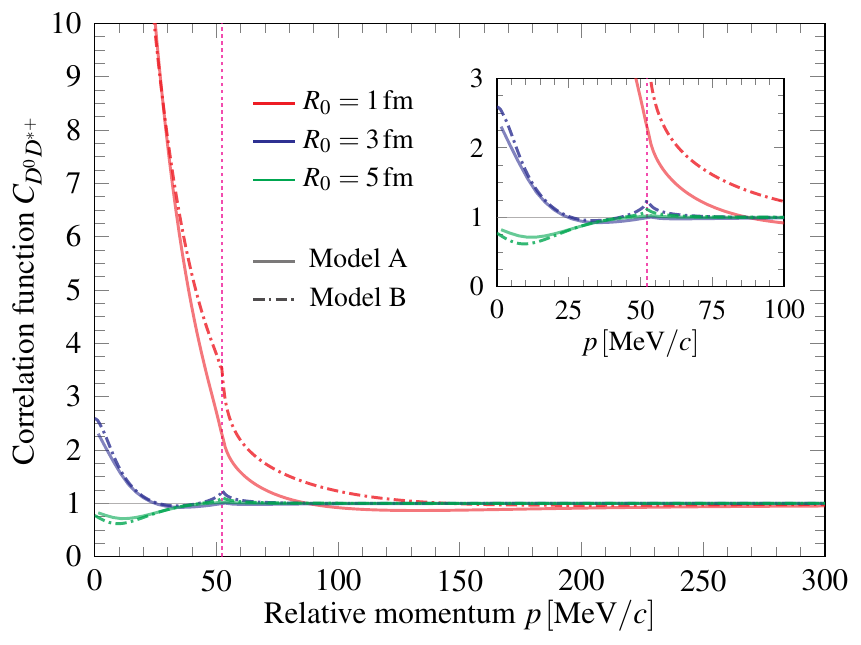}%
    \includegraphics[width=0.4\textwidth]{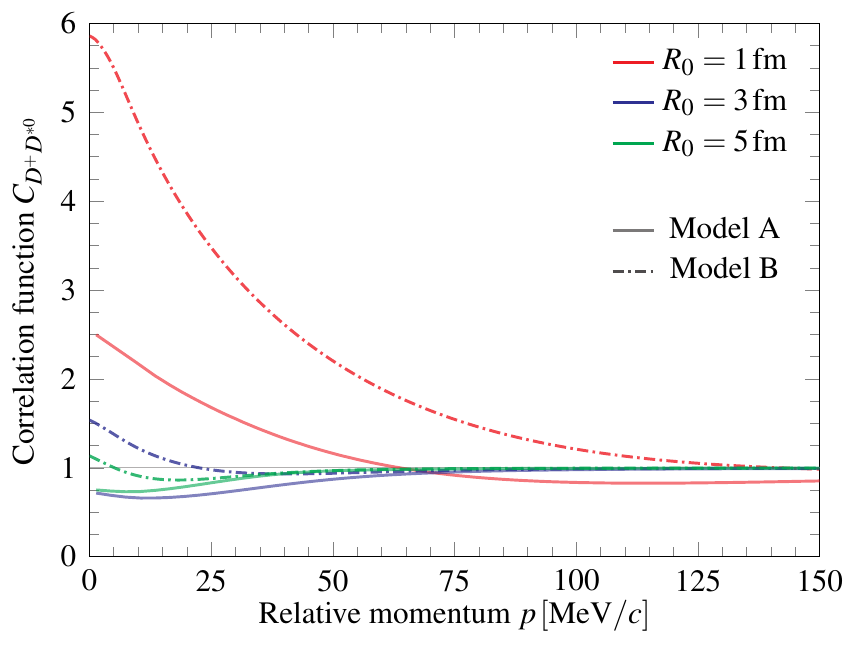}
    \caption{Correlation functions for the $D^0 D^{\ast+}$ (left) and $D^+ D^{\ast0}$ (right) pairs. The solid lines, denoted by ``Model A'', show the correlation functions computed as discussed in Sec.~\ref{sec:results}, using as an input the amplitudes derived in  Ref.~\cite{Feijoo21}. The dash-dotted line, denoted by  ``Model B'', represent the same correlation functions but calculated with the modifications discussed in Sec.~\ref{sec:compar}, using as an input the $T$-matrix from Ref.~\cite{Albaladejo22}. Red, blue, and green lines correspond to a source size $R_0=1$, $3$, and $5\,\text{fm}$, respectively. The vertical, magenta, dotted line on the left plot indicates  the $D^0 D^{\ast+}$ threshold.%
    \label{fig:ComparisonOtherAmplitude}}
\end{figure*}

In Sec.~\ref{sec:results}, the correlation functions for the $D^0 D^{\ast+}$ and $D^+ D^{\ast0}$ pairs (in the $T_{cc}^+$ channel) have been computed with the formalism discussed in Sec.~\ref{sec:tcc}. There, we have taken as an input the scattering matrix of Ref.~\cite{Feijoo21}, which provides a reasonable reproduction of the experimental mass and width of the $T_{cc}^+$ state, and thus constitutes a good baseline for our study. In this section, and in order to have an estimation of the theoretical uncertainties in our calculation, we employ instead the amplitudes calculated in Ref.~\cite{Albaladejo22}. In this latter work, the experimental LHCb $D^0 D^0 \pi^+$ spectrum where the $T_{cc}^+$ peak appears is fitted. The main difference of Ref.~\cite{Albaladejo22} with respect to the model of Ref.~\cite{Feijoo21} is that the interaction potential is not derived from the exchange of vector mesons dictated by the hidden gauge symmetry approach, but two independent constants appearing in a general-isospin symmetric potential are fitted to the experimental spectrum. Furthermore, the loop functions entering the $T$-matrix are not regulated with a sharp cutoff, but with a Gaussian one, and non-relativistic kinematics are explicitly used. Thus, in the alternative calculation of this section, we use the scattering amplitudes obtained in Ref.~\cite{Albaladejo22} and, for consistency, we employ non-relativistic kinematics and a Gaussian regulator in the computation of the quantities $\widetilde{G}_i(r;E)$.

The correlation functions obtained with this new scheme are displayed in Fig.~\ref{fig:ComparisonOtherAmplitude} with dash-dotted lines, and compared with those reported in Sec.~\ref{sec:results}, which are shown with solid lines. As can be seen, there is very good agreement for the $D^0 D^{\ast +}$ correlation function for the three sizes considered ($R_0=1$ (red), $3$ (blue), and $5\,\text{fm}$ (green)). This agreement is due to the similar $T_{cc}^+$ pole description found in Refs.~\cite{Feijoo21,Albaladejo22}. For the $D^+ D^{\ast 0}$ pair there are some discrepancies, but the trend is very similar in both cases and, more importantly, this channel is further away from the $T_{cc}^+$ pole than the $D^0 D^{\ast+}$ one, which is the most relevant one to obtain the $T_{cc}^+$ properties.

\section{Conclusions}
\label{sec:conclusions}

We have carried out a detailed study of the correlation function $C(p)$ for the $D^0 D^{*+}$ and $D^+D^{*0}$ channels that build up the $T_{cc}(3875)^+$ state, only a few hundred keV below the $D^0D^{*+}$ threshold. We find values of $C(p)$ positive and very large, around 30 for a source size $R=1$ fm, at the threshold for the $D^0D^{*+}$ channel and also positive and bigger than 1 for the  $D^+D^{*0}$ channel, but substantially smaller than for $D^0 D^{*+}$. Both correlation functions converge rapidly to unity for values of the relative momentum of the pair, $p$, around 200 MeV. We have performed several tests to see which are the terms that contribute most to the correlation function and found that all of them, interference of incoming wave with the re-scattered wave, and the squared of the re-scattered one, are important, particularly that of the observed channel. We modified slightly the formalism of Koonin--Pratt and showed that it is possible to factorize the scattering amplitude outside the integrals and only the range of the interaction has an effect in regularizing the loop function entering the formalism. Even then, we proved that with values of $R=1-5$ fm for the Gaussian source function, the range of the strong interaction becomes almost inoperative in the present case, once the proper scattering matrices are used. This is because the Bessel functions entering the formalism provide a stronger cut in the loop integration than the one provided by the range of the strong interaction. 
We conducted another important test which is that the substitution of the $T$-matrix by $T\propto(-1/a -ip)^{-1}$ in the effective range expansion, already provides an excellent approximation to the exact solution, although for the case of the $D^0D^{*+}$ channel, the consideration of the effective range improves the agreement.  This makes us reach the conclusion that one can only hope to accurately get the scattering length from these reactions, provided that from phenomenology we can get a hold on a reasonable value of the parameter of the Gaussian source function.
 Even then, the possibility of measuring correlation functions for pairs of particles that cannot be reached in scattering experiments, opens the door to find very valuable information on the interaction of hadrons. There is still a caveat for the case of coupled channels,
 since we found that the contribution of the non observed channel to the correlation function is not small. In other words, for the two channel case the correlations depend on $T_{11}, T_{12}, T_{22}$ and we only have two correlation functions, so we cannot determine the three magnitudes from two observables. Yet, by using the optical theorem one can relate $|T_{12}|^2$ with the imaginary part of $T_{22}$ and one can exploit this feature. In any case, the value of the correlation function could serve to test models rather than to build them. In the present case, the measurement of the correlation functions for the channels of the $T_{cc}$ would provide additional information to contrast what has been obtained from the analysis of the data on the $D^0D^0\pi^+$ spectrum observed in the LHCb experiment.   




\section*{Acknowledgments}
This work was supported by the Spanish Ministerio de Ciencia e Innovaci\'on (MICINN) and European FEDER funds under Contracts No.\,PID2020-112777GB-I00, and by Generalitat Valenciana under contract PROMETEO/2020/023. This project has received funding from the European Union Horizon 2020 research and innovation programme under the program H2020-INFRAIA-2018-1, grant agreement No.\,824093 of the STRONG-2020 project. M.\,A. and  A.\,F. are supported through Generalitat Valencia (GVA) Grants Nos.\,CIDEGENT/2020/002 and APOSTD-2021-112, respectively. M.\,A. and A.\,F. thank the warm support of ACVJLI. 





\end{document}